**Designing *de novo* TIM Barrels: Insights into Stabilization, Diversification, and Functionalization Strategies**

Julian Beck,[1] Sergio Romero-Romero[2*]

[1] Department of Biochemistry, University of Bayreuth, 95447 Bayreuth, Germany.

[2] Department of Biochemistry and Structural Biology, Instituto de Fisiología Celular, Universidad Nacional Autónoma de México, 04510 Mexico City, Mexico.

**\*Correspondence**

Sergio Romero-Romero: sromero@ifc.unam.mx

**ORCID identifiers**

Julian Beck: 0009-0007-3555-9890

Sergio Romero-Romero: 0000-0003-2144-7912

**Perspectives**

- **Highlight the importance of the field:** This manuscript highlights the TIM-barrel fold as a key model in protein design, emphasizing its versatility and challenges in functionalization. This review was undertaken to explore recent progress in TIM barrel design, identify key limitations in functionalization, and discuss emerging strategies, particularly AI-driven approaches, to overcome these challenges and expand the utility of TIM barrels in protein engineering.

- **Summary of the current thinking:** The review outlines advances in *de novo* TIM barrel design, highlighting challenges in functionalization and recent AI-driven strategies. It emphasizes the need for integrated approaches to optimize both scaffold stability and active site formation, paving the way for custom enzyme development.

- **Comment on future directions:** While significant advances have been made in *de novo* TIM barrel design, achieving enzymatic activity remains difficult due to the lack of natural-like active sites. Current strategies involve structural modifications and AI-driven approaches to tailor scaffolds for specific functions. These insights have important implications for human health and disease, as engineered TIM barrels could lead to novel enzymes for drug development, enzyme replacement therapies, diagnostics, and biocatalysis. By integrating computational and experimental methods, TIM barrels serve as a powerful template for custom enzyme design with potential biomedical and biotechnological applications.

**Abstract**

The TIM-barrel fold is one of the most versatile and ubiquitous protein folds in nature, hosting a wide variety of catalytic activities and functions while serving as a model system in protein biochemistry and engineering. This review explores its role as a key fold model in protein design, particularly in addressing challenges in stabilization and functionalization. We discuss historical and recent advances in *de novo* TIM barrel design from the landmark creation of sTIM11 to the development of the diversified variants, with a special focus on deepening our understanding of the determinants that modulate the sequence-structure-function relationships of this architecture. Also, we examine why the diversification of *de novo* TIM barrels towards functionalization remains a major challenge, given the absence of natural-like active site features. Current approaches have focused on incorporating structural extensions, modifying loops, and using cutting-edge AI-based strategies to create scaffolds with tailored characteristics. Despite significant advances, achieving enzymatically active *de novo* TIM barrels has been proven difficult, with only recent breakthroughs demonstrating functionalized designs. We discuss the limitations of stepwise functionalization approaches and support an integrated approach that simultaneously optimizes scaffold structure and active site shape, using both physical- and AI-driven methods. By combining computational and experimental insights, we highlight the TIM barrel as a powerful template for custom enzyme design and as a model system to explore the intersection of protein biochemistry, biophysics, and design.

***Designing the Foundation:*** **Why are TIM Barrels relevant to *de novo* protein design?**

Proteins are biological machines, carrying out a myriad of functions in life. The ability to design proteins with tailored properties or functions offers great opportunities in biochemistry, biotechnology, or biomedicine, from new therapeutic options to the development of better industrial enzymes. Herein, the structural stabilization of proteins and their functionalization represent two significant challenges within the context of protein design. While thermodynamic and kinetic stabilization allows proteins to exist stably under various conditions (Goldenzweig and Fleishman 2018; Romero Romero et al. 2018; Sanchez-Ruiz 2010; Pace and Grimsley 2001; Fersht et al. 1993; DeGrado et al. 1989), functionalization involves improving some specific biochemical functions and amino acid geometries onto proteins of interest (Pan and Kortemme 2021; Khakzad et al. 2023; Notin et al. 2024; Lechner et al. 2018). Tackling such issues requires a profound insight not only into the structure and dynamics of proteins but also a generally versatile scaffold amenable to successful design and engineering.

One of the most ubiquitous and versatile topologies in nature, the TIM-barrel fold, has emerged as a key model system in protein design (Romero-Romero et al. 2021b; Huang et al. 2016; Offredi et al. 2003; Figueroa et al. 2013; Tanaka et al. 1994; Beauregard et al. 1991). Composed of eight β-strands that alternate with eight α-helices, the TIM barrel possesses a very symmetric and modular architecture (Nagano et al. 2002a; Wierenga 2001; Höcker et al. 2004; Lang et al. 2000). This fold is remarkably versatile, supporting a wide range of catalytic functions while allowing for extensive variations in sequence and structure with the conservation of its core topology. Its ancient evolutionary origins further support its importance, being the TIM barrel a fold that has diversified across billions of years of evolutionary history, hosting multiple types of activities and functions (Nagano et al. 2002a; Copley and Bork 2000; Henn-Sax et al. 2001; Anantharaman et al. 2003; Goldman et al. 2016; Sterner and Höcker 2005).

Its size and complexity render the TIM barrel particularly appealing for protein design, a compromise between structural simplicity and architectural sophistication. This topology provides a well-defined framework in which protein design rules can be applied, including modular design strategies and principles derived from sequence-folding-structure relationships (Koga et al. 2012; Koga and Koga 2019; Nagarajan et al. 2015; Kadumuri and Vadrevu 2018; Romero-Romero et al. 2018; Quezada et al. 2018; Romero-Romero et al. 2015). Besides, stability, dynamics, and modularity make it a suitable system for testing the effects of changes in sequence on protein stability and structure, thus offering an ideal architecture to study sequence-structure-function relationships.

In this mini-review, we discuss the role of *de novo* TIM barrels in protein design and engineering studies. We will also discuss how they serve as a model system in addressing important challenges: stabilization via computational and experimental approaches, and diversification toward functionalization and tailor-enzyme design. By examining historic advances in context with more recent ones, we show how the unique properties of TIM barrels —ancient, versatile, designable, and amenable to investigations into sequence-structure-function relationships— have made them indispensable architectures for pushing the boundaries of protein science.

***Designing for Stability*: Unraveling the Stability Landscape of *de novo* TIM Barrels**

Protein stability is a crucial aspect to consider, as it directly affects functionality and applicability in biochemistry, biotechnology, and medicine. Understanding and manipulating protein stability may not only provide key insights into the molecular determinants of the protein sequence-structure-function relationship but also increase our knowledge about the protein design rules. Within this context, TIM barrels have been a model system to study how changes in the amino acid sequence influence protein stability and a long target of *de novo* protein design efforts. Figure 1 and Table 1 summarize the main studies and approaches in this field, as well as efforts toward diversification and functionalization, which are described in the next section.

In 2016, Huang and co-authors reported a four-fold symmetric *de novo* TIM-barrel protein (Huang et al. 2016), constituting an important landmark in this field so far, due to the historical difficulty of obtaining accurate and stable structures of this kind. This work used the Rosetta software to follow a bottom-up approach based on geometric and chemical principles. Also, it provided design rules that included specific side chain-backbone hydrogen-bonding interactions that were critical for sustaining strand register across repeat units. The experimentally validated design, sTIM11, was a thermostable protein of 184 residues that demonstrated proper secondary and tertiary structure element arrangements as indicated in the Rosetta model. Importantly, sTIM11 exhibited a structure compatible with the intended TIM barrel topology but with a sequence not similar to natural TIM barrels.

Using sTIM11 as a scaffold, Romero-Romero et al. designed a large family of *de novo* TIM barrels, DeNovoTIMs, using a computational fixed-backbone and modular approach with a focus on improving stability by enhancing hydrophobic packing based on the first validated *de novo* TIM barrel, sTIM11 (Romero-Romero et al. 2021a). The DeNovoTIMs exhibited a wide range of both thermal and conformational stabilities, extending their applications in biochemistry, as will be further discussed in the next section. Furthermore, thermodynamic analyses suggested that large non-additive effects modulate the stability effects, pointing out that the stability of DeNovoTIMs is tunable by epistatic effects, i.e., mutations across different regions of the barrel.

In follow-up work, Kordes et al. tested the impact of introducing a salt bridge cluster on the structural and biophysical properties of three previously designed DeNovoTIMs (Kordes et al. 2022). While the salt-bridge variants retained similar thermostabilities compared to the parental proteins, they showed differences in conformational stabilities at room temperature and different crystallization tendencies. Besides, the structural analyses showed that the geometries of the salt bridges across the proteins varied, directly influencing their stability and crystallization properties. This work underlined the complexity of the prediction of salt bridge interactions, suggesting that the design of such clusters in *de novo* proteins is an open field for further investigations.

Recently, Koch and colleagues studied the effects of the stepwise introduction of stabilizing mutations across the four TIM-barrel quarters on the thermal and conformational stability properties (Koch et al. 2024). The study was built upon the foundation of earlier designs, particularly those of members of the DeNovoTIMs collection. They introduced mutations in a quarter-wise manner from DeNovoTIM0, the non-stabilized design, to DeNovoTIM6, a stabilized barrel, and observed non-linear and non-additive effects of stability, meaning that the number and location of mutations strongly influence overall protein stability. Accordingly, equivalent mutations in different structural positions of the protein also altered the stability, emphasizing the importance of effective packing and hydrophobic interactions to be achieved in such a closed architecture as the TIM-barrel fold. Results showed the complexity of modulating protein stability by sequence changes and gave valuable insights into protein design and engineering.

Overall, this stability landscape navigation by modifying the protein sequence has not only increased our knowledge about how stability can be fine-tuned by protein design but also expanded the repertoire of synthetic proteins. It laid the foundational principles that could enable the design of other complex protein architectures and, therefore, opened up possibilities for customized enzyme design.

***Designing for Functionality:*** **Diversification and Optimization Strategies for Tailored Purposes**

The design of the first *de novo* TIM barrel in 2016 (Huang et al. 2016) raised hopes that tailor-made enzymes were within reach. However, despite the passing time and recent advances in protein design based on the introduced AI tools, the achievements in functionalization do not yet completely fulfill the high expectations. A comparison of *de* *novo* TIM barrels to natural sheds light on the underlying challenges of functionalization. Natural TIM barrels form pockets using extended loops, secondary structure elements, or even additional domains, often anchoring critical catalytic residues within these extensions (Nagano et al. 2002b; Sterner and Höcker 2005; Romero-Romero et al. 2021b). Due to their highly idealized structure, such features are absent in *de novo* TIM barrels. To overcome these limitations, multiple efforts were made to diversify the provided scaffold to move *de novo* TIM barrels toward diversification and functionalization.

As a key model system in protein design and a challenging design target, the TIM-barrel fold was used by multiple groups to validate their design pipelines. Anand and colleagues tested the capability of a deep neural network model for sequence design, achieving multiple well-behaved proteins (hereafter referred to as F-barrels for simplicity) and adding several *de novo* TIM barrels to the existing repertoire (Anand et al. 2022). In another work, Goverde and colleagues developed a deep learning pipeline to design complex folds and soluble analogs of integral membrane proteins (Goverde et al. 2024). This pipeline resulted in new *de novo* TIM barrels with diversified sequences even breaking the established four-fold symmetry. Additional *de novo* TIM barrels were generated by Watson and colleagues to demonstrate that their generative diffusion algorithm (RFdiffusion) can be conditioned towards specific folds (Watson et al. 2023). The proteins generated exhibit the expected circular dichroism (CD) and a high thermostability, but a detailed structural characterization was not performed. Despite another focus in all these studies, they diversified *de novo* TIM barrels to the existing repertoire either on the structural or sequence level.

To achieve a different structural starting point compared to the circular sTIM11 and its descendants, Chu *et al.* designed a *de novo* TIM barrel with an ovoid shape (Chu et al. 2022). This ovoid curvature aimed to accommodate residue combinations and networks within the core that might not be feasible within a circular TIM barrel. For the design of an ovoid TIM barrel, they constructed a blueprint for an ovoid topology of the inner β-barrel guided by a two-fold symmetry, varying the curvature and structural parameters such as the shear number. An autoregressive sampling strategy was used to sequentially build each secondary structure element to achieve optimal α-helix and loop lengths to support the ovoid curvature, and sequence design was performed via an iterative enrichment protocol with Rosetta (Ren et al. 2022; Huang et al. 2011). The resulting ovoid TIM barrels showcased high thermodynamic stability, and a solved crystal structure confirmed a high similarity to the computational model and the intended ovoid geometry. Notably, additional features favorable for downstream functionalization include βα-loops, which are free of any critical hydrogen bonding, and a high tolerance of the inner core to polar or charged residues eventually necessary for function. While these features make the ovoid TIM barrels a good alternative starting point, some critical limitations remain, like the absence of pockets, extended loops, or structural extensions to accommodate functional residues making downstream functionalization still a demanding task.

To move towards functionalization with another strategy, multiple secondary structure elements have been added on top of sTIM11 and its descendants, aiming to achieve additional surface area and potential catalytic pockets. For instance, Wiese *et al.* incorporated a small α-helix into the scaffold, a common motif in natural TIM barrels often involved in phosphate binding (Wiese et al. 2021). Their approach involved Rosetta *ab initio* structure prediction of the intended α-helix, its attachment to the scaffold, and Rosetta optimization. However, the crystal structure showed significant deviation from the intended design, with the desired α-helix adopting a $3_{10}$ configuration. This discrepancy, combined with the small size of the added motif, limited the potential for functionalization.

In another study, Kordes *et al.* introduced substantially larger secondary structure elements (helix-loop-helix motifs) to significantly expand the surface area and establish a pocket above the barrel, generating so-called αTIMs (Kordes et al. 2023). Their workflow for inserting a single helix-loop-helix motif involved rational design, Rosetta *ab initio* structure prediction, attachment to the scaffold, and optimization of the motif and its transition region into the barrel using Rosetta. As initial experiments indicated the successful extension, the scaffold's four-fold symmetry was leveraged to duplicate the motif in another half of the barrel, enhancing the surface area further and potentially forming a pocket between the extensions. While no experimental structure was obtained, AlphaFold2 predictions supported the Rosetta models, and PUResNET (Kandel et al. 2021) predicted the formation of a pocket above the inner β-barrel, achieving the design objectives.

Notably, both extended TIM barrels relied on Rosetta *ab initio* structure prediction rather than the now-established AlphaFold2 (Jumper et al. 2021). This is easily explained by the fact that both studies began long before AlphaFold2's release, underscoring the lengthy timeline of efforts to diversify *de novo* TIM barrels. In contrast, a more recent study by Beck *et al.* used AlphaFold2 alongside other AI tools for sequence and structure diversification (Beck et al. 2024). Unlike earlier approaches based on rational design, their method relied on constrained hallucination, which is theoretically capable of generating diverse structural elements (Anishchenko et al. 2021; Wang et al. 2022). Despite this potential, Beck *et al.* obtained exclusively helical extensions closely resembling the helix-loop-helix motif previously introduced by Kordes *et al*. A notable advancement was the successful addition of an extra extension, increasing surface area and enhancing potential pocket interactions. However, crystal structures and SAXS data indicated substantial flexibility in the extensions, making any functionalization of the predicted pocket highly challenging.

One additional downside of these stepwise introductions of pockets for downstream functionalization is that they are introduced as generalizable pockets without considering a certain enzyme or binding activity. This might be advantageous initially, but introducing function, specifically enzymatic activity, requires exact control over the geometry of the catalytic residues and their surroundings. The chances that a premade pocket can accommodate the desired reaction and catalyze it with desirable efficiency are unlikely, making the strategy of stepwise diversification towards functionality uncertain.

Despite these challenges, Caldwell *et al.* demonstrated an outstanding example through the successful stepwise introduction of a binding functionality into a highly modified *de novo* TIM barrel (Caldwell et al. 2020). The initial step involved further optimization and reversal of the circular permutation in the already optimized DeNovoTIMs, resulting in DeNovoTIM15 with an additional βα-loop compared to the existing DeNovoTIMs. Additionally, they utilized the barrel's four-fold symmetry by splitting it into two halves, achieving a complete TIM barrel through homodimerization. By fusing a monomer of a de novo-designed ferredoxin (Lin et al. 2017), they generated a highly stable homodimeric fusion protein, TIM-FD (TFD), which forms a substantial internal cavity above the barrel, between the domains. To functionalize the designed cavity, they introduced two glutamates to build up a metal coordination site for large trivalent cations (TFD-EE variant). They showcased lanthanide binding with a high affinity via spectroscopic measurements and a high-resolution crystal structure. With the binding of lanthanides and the availability of designable residues in the cavity for substrate recognition, the protein shows potential for downstream enzyme design. Nonetheless, it still faces the drawbacks of a premade pocket, which requires further modifications to achieve the necessary customizability for tailor-made enzymes.

Building upon this design, Klein et al. recently repurposed the TFD scaffold to create a *de novo* photoenzyme for lanthanide-mediated photoredox catalysis (PhotoLanZyme) (Klein et al. 2024). The design incorporated a tetraglutamate $Ce^{3+}$ binding site in the internal cavity, enabling light-triggered ligand-to-metal charge transfer that initiates selective C–C bond cleavage in 1,2-diols. To improve metal specificity and reduce unspecific terbium binding to the protein surface, the authors identified exposed acidic residues, mutating six aspartates and two glutamates to neutral residues, yielding PhotoLanZyme version 1.4. Additionally, the engineered enzyme retained function when expressed on the surface of *E. coli*, enabling photobiocatalysis in whole cells. This work exemplifies how a well-characterized, modular *de novo* scaffold can be progressively adapted toward increasingly complex and abiological catalytic functions.

In a follow-up study, Egea and colleagues aimed to alter the metal specificity toward divalent first-row transition metal ions by introducing the mutation E154H into each monomer of the scaffold (TFD-EH) (Egea et al. 2025) (Table 1). The introduced mutation enabled binding of divalent metals, as demonstrated by crystal structures of complexes with $Cu^{2+}$ and $Ni^{2+}$, in which even two ions were bound. To further stabilize a dinuclear binding mode, an additional coordinating residue (T87E) was introduced. Crystal structures with $Zn^{2+}$ and $Co^{2+}$ demonstrated the participation of this residue and revealed a different binding coordination compared to before. As the TFD scaffold contains tri-glycine linkers, the authors also investigated the flexibility of their variants, revealing that the TFD scaffold is not rigid and that its flexible linkers enable dynamic motions. Furthermore, this analysis showed that TFD-EE adopts a conformational equilibrium between two states. To shift the equilibrium toward the active state, the scaffold was optimized using ProteinMPNN. The resulting protein, TFD-EE MPNN, formed only a single dimeric species and exhibited a 10-fold increase in catalytic efficiency compared to the original scaffold (Table 1). This study highlights the possible plasticity of *de novo* scaffolds and demonstrates the potential of multi-domain architectures with flexible linkers to enable conformational dynamics in hyperstable *de novo* proteins

To harness the full enzymatic potential of *de novo* TIM barrels, a paradigm shift in their functionalization is necessary. Achieving the required accuracy for a desired enzymatic reaction with sufficient efficiency through a stepwise addition of extensions and functional sites is exceedingly difficult (Dawson et al. 2019). Only recently Beck and colleagues achieved an enzymatic active *de novo* TIM barrel based on the simultaneous introduction of an extension to form an active site for a specific reaction (Beck et al. 2025).

With their so-called CANVAS approach (Customizing Amino-acid Networks for Virtual Active-site Scaffolding), which includes physics-based software like Triad (Lee et al. 2023) and AI-based tools like RFdiffusion (Watson et al. 2023), they introduced a Kemp eliminase activity —a benchmark reaction in computational enzyme design for carbon-based proton transfer— into a circular-permutated *de novo* TIM barrel.

One active variant showcased a comparable efficiency to traditionally designed Kemp eliminases with transition state placements in existing scaffolds (Röthlisberger et al. 2008), despite the additional challenge of building up structural extensions for the active site. Nonetheless, in comparison to recently achieved catalytic efficiencies in *de novo* enzyme design with natural and entirely newly created scaffolds (Lauko et al. 2025; Yeh et al. 2023; Braun et al. 2024; Romero-Romero et al. 2024), the achieved efficiency is lower and far away from being the desired outcome of *de novo* enzyme design or harnessing the full potential of the TIM-barrel fold. Nonetheless, with these recent advances in enzyme design, alongside progress in general protein design and the emergence of AI-based tools (Vázquez Torres et al. 2025; Glögl et al. 2024; Sumida et al. 2024; Pillai et al. 2024; Ingraham et al. 2023; Schneuing et al. 2024; Pacesa et al. 2024), we are optimistic that increasingly complex and efficient enzymatic TIM barrels and the exploration of their full enzymatic potential are on the horizon.

*Designing Beyond*: **Concluding Perspectives and Challenges in the Field**

The TIM-barrel architecture represents an optimal topology to apply protein design rules and increase our understanding of protein structure and function while unlocking new applications in biochemistry and synthetic biology. As discussed in this review, the successful design of TIM-barrel proteins highlights the importance of coupling computational techniques with experimental validation. Recent advances in computational protein design, particularly through the application of physics-based methods and deep learning algorithms, have led to the creation of novel TIM-barrel members with unprecedented stability, sequence diversity and, recently, catalytic activity. However, several obstacles remain in the field.

One of the major challenges is extending the functional profile of the designed and engineered proteins, as naturally occurring TIM barrels can bind complex ligands or catalyze complex reactions. It calls for tight control of binding sites and active site geometries to introduce novel functions or cofactors. Although expanding the TIM barrel topology by adding other functional domains or motifs can lead to novel scaffolds with improved traits, this requires a better understanding of how different structural elements can be properly fused without loss of stability or function. Eventually, it should even be considered to shift from fusing existing TIM barrels to designing an entirely new TIM barrel in the context of the desired function, as this could lead to a superior interplay between extensions and core, enabling a broader range of possible topologies for extensions.

Furthermore, protein dynamics is crucial for enzymatic function, yet most design approaches target static structures. Capturing and designing conformational flexibility remains a significant task since most traditional structure prediction algorithms predict single, rigid conformations. To design highly functional TIM-barrel enzymes, it is crucial to validate and apply new methods that incorporate dynamics explicitly, either through molecular dynamics simulations or AI-based approaches.

Having achieved all this insight in the TIM-barrel fold, it is also worth exploring how such design principles could be extended to other protein topologies. While TIM barrels provide a robust and adaptable scaffold, certain catalytic activities or binding specificities might be better suited by other architectures, i.e., α/β-hydrolases, Rossmann folds, or even novel *de novo* structures. By the application of computational and experimental strategies developed for TIM-barrel design, it is now feasible to generate entirely new protein folds specific to functional requirements. This expansion beyond TIM barrels could lead to new enzymatic activities and increase the number of de novo-designed proteins available for new applications.

Finally, as has been discussed, a promising direction for the field is the integration of physics-based and AI-driven approaches. While physics-based design provides mechanistic insight into the design process and energy calculation, AI methods offer effective sequence and structure optimization. Assessing the performance of these combined approaches along with addressing the challenges in the field might lead to the generation of more robust and efficient proteins, reducing costs and time for the computational and experimental characterization, and allowing the design of more complex artificial proteins with tailored functionalities.

**Acknowledgments**

We thank Daniel Ríos Barrera for his proofreading and valuable suggestions to improve the manuscript.

**Funding**

J. B. acknowledges support from the Elite Network of Bavaria and its "Biological Physics" study program. This work was supported by CONAHCYT (grant: CBF2023-2024-271) and UNAM-PAPIIT (grant: IA203925).

**Competing interests**

The authors declare no conflicts of interest in relation to the contents of this article.

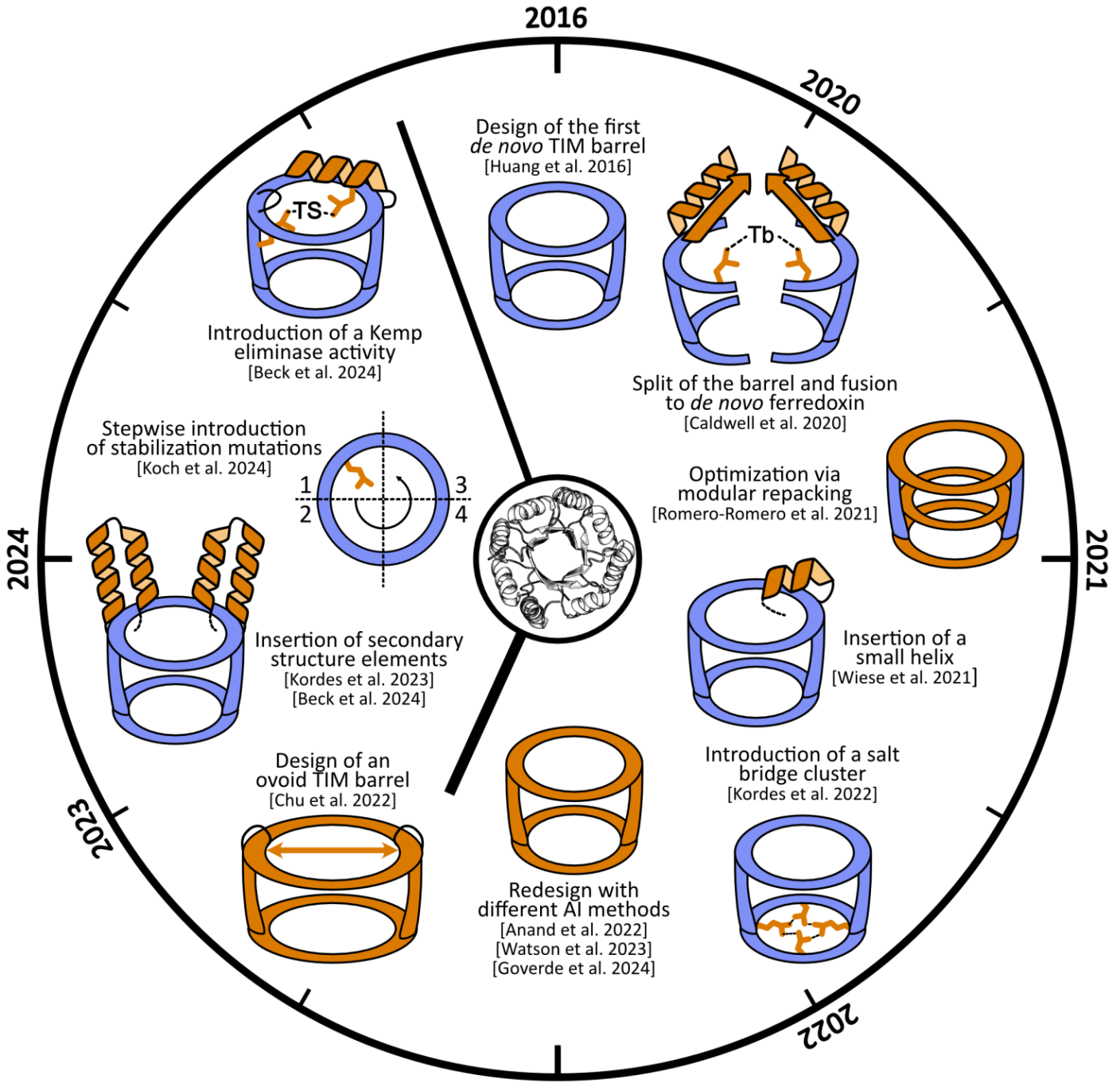

**Figure 1. Schematic overview of all studies involving *de novo* TIM barrels.** The diagram begins with sTIM11, the first *de novo* TIM barrel, positioned at 12 o'clock, with its crystal structure as a white cartoon representation in the center (PDB: 5BVL). In the schematic, the TIM-barrel fold is simplified as a barrel. Blue indicates major preservation of the starting scaffold, while orange highlights the main objective or modification of each study. The timeline progresses clockwise, with the years on the outer edge marking the publication year of each reference or the first reference within a similar topic.

**Table 1.** Summary of *de novo* designed TIM barrels to date and their design strategies.

| Protein/Family name | PDB IDs of related structures | Approach classification | Brief description and design strategy | Reference |
|---|---|---|---|---|
| **Aim: Stabilization** | | | | |
| sTIM11 | 5BVL | Physics-based | The first structurally validated *de novo* TIM barrel, designed using Rosetta to build a 4-fold symmetric scaffold | (Huang et al. 2016) |
| sTIM11noCys | 6YQY | Physics-based | Cysteine-free variant from sTIM11; this modification prevents disulfide bond formation, enhancing scaffold applicability | (Romero-Romero et al. 2021a) |
| DeNovoTIMs | 6Z2I (DeNovoTIM6), 6YQX (DeNovoTIM13) | Physics-based | A series of *de novo* TIM barrels with diverse thermal and conformational stabilities, designed using Rosetta with a modular approach to enhance hydrophobic core packing and improve sTIM11 stability | (Romero-Romero et al. 2021a) |
| DeNovoTIM15 | 6WVS | Physics-based | Circularly permutated variant of DeNovoTIM13, designed using RosettaRemodel to revert the circular permutation of the original sTIM11 | (Caldwell et al. 2020) |
| DeNovoTIMs-SB | 7OSU and 7OT7 (sTIM11noCys-SB), 7OSV and 7OT8 (DeNovoTIM6-SB), 7P12 (DeNovoTIM13-SB) | Physics-based | DeNovoTIM variants designed using Rosetta based on natural TIM barrels, incorporating central salt bridge clusters to enhance structural stability and facilitate crystallization | (Kordes et al. 2022) |
| DeNovoTIMs-quarters | No structures available | Physics-based | TIM barrel proteins with quarter-turn symmetry mutations, designed to explore the role of non-additive effects in protein stability | (Koch et al. 2024) |

| | | | Aim: Diversification & Functionalization | |
|---|---|---|---|---|
| TIM-FDs | 6WXO (TFD-HE), 6WXP (TFD-EE), 6ZV9 (TFD-EE N6W) | Physics-based | Variant with a large internal cavity featuring a lanthanide-binding site for metal coordination, created by fusing two domains: a *de novo* designed dimeric ferredoxin fold and DeNovoTIM15 | (Caldwell et al. 2020) |
| sTIM11_helix3 | 7A8S | Physics-based | Extension of the sTIM11noCys variant using Rosetta, incorporating a small helix at the top of the barrel for structural enhancement | (Wiese et al. 2021) |
| OvoidTIM3 | 7UEK | Physics-based | A two-fold symmetrical design of an ovoid-shaped TIM barrel, with structural loops generated via RosettaRemodel | (Chu et al. 2022) |
| F-barrels | 7MCC (F2C), 7SMJ (F2N), 7MCD (F15C) | AI-based | Variants with sequences derived from a learned potential using a deep neural network model | (Anand et al. 2022) |
| αTIMs (αTIM2 and αTIM2-2) | No structures available | Physics-based | Rationally designed coiled coils introduced into the top region of sTIM11noCys using RosettaRemodel | (Kordes et al. 2023) |
| TIM_barrel_6 | No structure available | AI-based | Non-idealized TIM barrel protein designed with RFdiffusion for structural and sequence variation | (Watson et al. 2023) |
| TBF_24 | 8OYS | AI-based | Redesign of TIM barrel proteins through an inversion of AlphaFold2 coupled with ProteinMPNN to mimic the topology of natural membrane proteins | (Goverde et al. 2024) |
| HalluTIMs | 8R8N (HalluTIM2-2), 8R8O (HalluTIM3-1) | AI-based | Structural extensions of DeNovoTIMs with multiple α-helical hairpins, designed via constrained hallucination | (Beck et al. 2024) |
| PhotoLanZymes + TFD variants | 9QUC (TFD-EH), 9QUD (TFD-EH $Cu^{2+}$), 9QUI (TFD-EH $Ni^{2+}$), 9QUL (TFD-EH T87E $Zn^{2+}$), 9QUO (TFD-EH T87E $Co^{2+}$), 9QUP (TFD-EE MPNN) | Structure-based engineering | *De novo* designed photoenzymes (cerium-dependent enzyme) based on a previously designed TIM-barrel scaffold (TIM-FD), enabling C–C bond cleavage of 1,2-diols in aqueous solution via photoredox catalysis. Rational mutations modified metal-binding specificity and conformational behavior. AI-guided stabilization of the active conformation resulted in an improved enzyme. | (Klein et al. 2024) and (Egea et al. 2025) |
| KempTIMs | 9QKX | Physics-based + AI-based | Enzymatically active *de novo* TIM barrel (Kemp eliminase). KempTIMs were designed using the CANVAS computational workflow, incorporating AI tools for structural extensions and the physic-based software Triad for transition state placement and pocket design. | (Beck et al. 2025) |